%% file: main_arxiv.tex
\documentclass[10pt,journal,twocolumn]{IEEEtran}
\PassOptionsToPackage{table}{xcolor}
\input{common}
\usepackage{acronym}
\input{acronym}
\usepackage{stfloats}
\usepackage{tikz}
\usetikzlibrary{matrix,fit,backgrounds}

\newsavebox{\FDAbgbox}
\newcommand{\FDABackground}[1]{%
  \definecolor{panelbg}{HTML}{FFFEFC}   %
  \sbox{\FDAbgbox}{#1}%
  \begin{tikzpicture}
    \node[inner sep=0pt] (FDAbgcontent) {\usebox{\FDAbgbox}};
    \begin{pgfonlayer}{background}
      \node[fit=(FDAbgcontent),rounded corners=5pt,fill=black!3,
        draw=black!20,line width=0.4pt,inner sep=6pt] {};
    \end{pgfonlayer}
  \end{tikzpicture}%
}

\begin{document}

\title{Beamfocusing Beyond the Near Field: \\ Frequency Diverse Arrays for OFDM-ISAC Systems}

\author{Arda Akyazi$^\star$, Ahmed Hussain$^\star$, Asmaa Abdallah, Ahmed M. Eltawil, and Abdulkadir Celik

\thanks{$^\star$Arda Akyazi and Ahmed Hussain contributed equally.}
}
\maketitle
\begin{abstract}
Traditional phased arrays (PAs) achieve near-field (NF) beamfocusing through spherical wavefront curvature, but lose this range selectivity in the far field (FF). Frequency diverse arrays (FDAs) instead introduce frequency offsets across antenna elements. These offsets synthesize a range-dependent beam whose beamdepth is governed by the aggregate frequency-offset span rather than the physical aperture, thereby extending finite
beamdepth into the FF. Although FDAs have been extensively studied for radar, their broader potential for wireless communications remains largely unexplored. Accordingly, this article examines how this capability can support wireless communications and integrated sensing and communication (ISAC). We compare PA and FDA focusing mechanisms, explain FDA coupling and periodicities, and assess coexistence with orthogonal frequency-division multiplexing (OFDM). Integer-multiple offsets preserve alignment with the OFDM subcarrier grid and avoid offset-induced intercarrier interference under ideal synchronization, but require suitable symbol mapping and receiver processing. We also distinguish FDA beamdepth from waveform-based radar range resolution and discuss their complementary roles in sensing. For the configurations studied, FDAs offer no additional single-user spatial degrees of freedom. Interestingly, their principal benefit is multiuser separation through location division multiple access (LDMA). Numerical case studies under an idealized FDA channel model show that FDA-LDMA achieves higher sum-rates than PA-based schemes for users with small angular separation in the FF, even after accounting for the bandwidth reserved for frequency offsets.
\end{abstract}

\section*{\textbf{Introduction}}
\IEEEPARstart{B}{eamforming} with traditional \acp{PA} has long been the cornerstone of directional wireless transmission that relies on plane-wave propagation to steer energy toward a specific angular direction. However, as physical apertures enlarge from massive to gigantic arrays, and operating frequencies extend into the \ac{mmWave} and \ac{THz} bands, the Rayleigh distance can span hundreds of meters. Consequently, communication may transition to radiating \ac{NF}, wherein the incident wavefront is spherical rather than planar \cite{10934779}. By matching the quadratic phase curvature across the aperture, the \ac{PA} concentrates energy at a specific location in the \ac{NF} region. This fundamental shift from angular beamsteering to beamfocusing endows the array with a finite beamdepth, thereby enabling a new multiplexing dimension such as \ac{LDMA}~\cite{10123941}. However, the distance-dependent beamfocusing with a \ac{PA} relies on wavefront curvature and is therefore confined to the \ac{NF}, limiting its ability to provide distance-domain discrimination in the \ac{FF}.

Unlike the \acp{PA}, the \ac{FDA} provides range selectivity even in the \ac{FF}~\cite{11162170} by employing small frequency shifts across its antenna elements. Originally proposed for radar~\cite{1631800}, the \ac{FDA} has been extensively developed for range-dependent clutter suppression \cite{11431119} and secure physical-layer transmission \cite{10068433}. Moreover, \acp{FDA} differ from frequency-scanning arrays, which achieve beam steering through time-varying frequencies applied uniformly across all spatial channels. Architecturally, a \ac{PA} requires a single \ac{LO} to provide the same carrier frequency to all elements, whereas an \ac{FDA} requires additional mixer circuitry to transmit signals with different carrier frequencies. In a fully digital architecture, these offsets can be generated directly in the digital baseband without requiring separate analog circuitry.

{Gigantic \acp{PA} exploit wavefront curvature in the \ac{NF} to enable distance-domain multiplexing. In contrast, \acp{FDA} achieve beamfocusing in the \ac{FF} by exploiting frequency offsets rather than relying on large antenna apertures. Although \acp{FDA} have been investigated for sensing and secure communications, their potential for spatial multiplexing remains largely unexplored. This motivates a unified comparison of \acp{PA} and \acp{FDA} across both the \ac{NF} and \ac{FF}. In this position paper, we first contrast their underlying beamfocusing mechanisms and examine the range--angle coupling and periodicities of the \ac{FDA} beam pattern. We then investigate \ac{FDA}--\ac{OFDM} architectures and their trade-offs in orthogonality, array gain, and bandwidth utilization, and discuss how \ac{FDA} range selectivity complements waveform-based sensing in \ac{ISAC}. Finally, we compare beamwidth, beamdepth, and single-user \ac{EDoF}, and evaluate multiuser sum-rates while accounting for the bandwidth consumed by frequency offsets. This comparison identifies the conditions under which \acp{FDA} offer a net multiplexing gain and highlights future research directions in practical architectures, receiver processing, resource allocation, and physical-layer security.}

\section*{\textbf{FDA Fundamentals}}
\label{sec:fundamentals}
\subsection*{\hspace{7 pt} \textbf{Two Ways to Focus a Beam}}
\label{sec:focus}
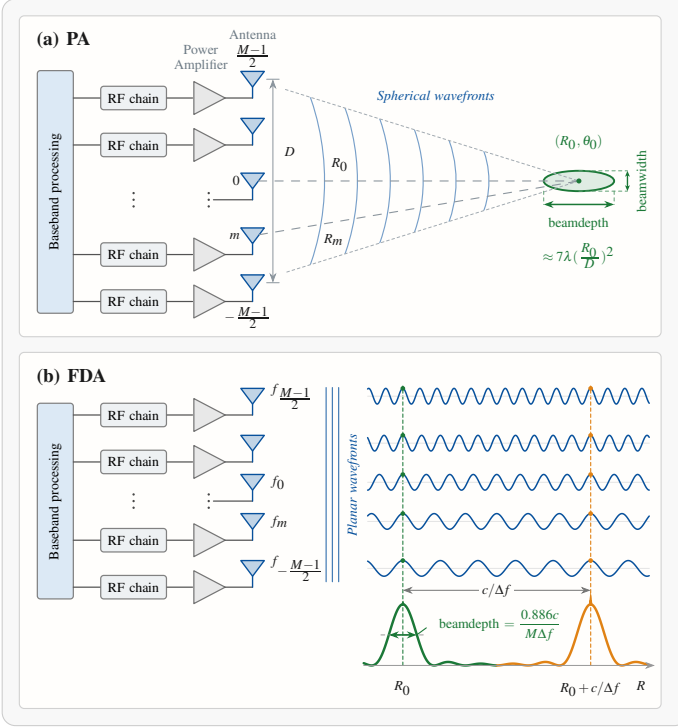
\begin{figure}[!t]\centering
\FDABackground{\resizebox{\dimexpr 0.5 \textwidth-12.4pt\relax}{!}{\input{fig1_combined}}}
\caption{Architecture and focusing mechanism. (a) \ac{PA} transmitter focuses by matching
\emph{wavefront curvature}, (b) \ac{FDA} transmitter focuses by \emph{aligning carrier phases} on a planar wavefront.}
\label{fig:arch}
\end{figure}

Two ways of beamfocusing are contrasted side by side in Fig.~\ref{fig:arch}(a) and (b).\footnote{We kindly refer readers to \url{https://akadircelik.com/demos/fda-ofdm-isac/} for an interactive and accessible version of the figures.}
An \ac{NF}-\ac{PA} leverages the array aperture to focus a beam. Its elements radiate at the same carrier frequency $f_c$ but with a progressive quadratic phase shift, causing the signals from different elements to interfere constructively at a prescribed focal point within the \ac{NF} region. The resulting beam has a finite beamdepth that can be approximated as $7\lambda(R_0/D)^2$~\cite{ebrd}, where $\lambda$ is the wavelength, $R_0$ is the focal distance, and $D$ is the maximum physical dimension of the array, such as the diagonal of a planar array. As $R_0$ increases, the wavefront curvature across a fixed aperture becomes weaker and the beamdepth broadens. Effective beamfocusing with a \ac{PA} is therefore confined to a fraction of the Rayleigh distance, and extending the focusing range requires a larger aperture.

An \ac{FDA}, by contrast, creates range selectivity through frequency-dependent phase accumulation rather than wavefront curvature. In a standard \ac{FDA}, adjacent antenna elements transmit at frequencies separated by a fixed increment $\Delta f$, introducing a range-dependent phase difference across the array. By appropriately configuring the initial phases, the signals from all elements can be aligned at a desired focal range, resulting in constructive interference. Away from this range, the accumulated phase differences cause the signals to become progressively misaligned, reducing the coherent array gain. Because this frequency-dependent phase term does not rely on wavefront curvature, it provides range selectivity in both the \ac{NF} and the \ac{FF}, irrespective of the Rayleigh boundary. In the \ac{FF}, where curvature-based focusing vanishes, the \ac{FDA} maintains a
beamdepth of $0.886c/(M\Delta f)$, governed by the aggregate frequency-offset span rather than by the physical aperture $D$. The range focus is periodic, however, and equivalent focal ranges recur at intervals of $c/\Delta f$.

\subsection*{\hspace{7 pt}  \textbf{Periodicities and Coupling}}

To expose the resulting periodicities and coupling, consider an $M$-element \ac{ULA} at the \ac{BS} where its elements are indexed as $m=0,\cdots,M-1$. The $m$-th antenna element transmits at an offset frequency ${f_m=f_c+\Delta f_m}$. The frequency offset $\Delta f_m$ is a design parameter of \ac{FDA}. For a standard linear \ac{FDA}, $\Delta f_m=m\Delta f$. A single-antenna \ac{UE} is located at range $R$ and angle $\theta$ relative to array broadside. Under a narrowband Fresnel approximation, the received phase contribution from element $m$ is given by~\cite{sajid2026}

\begin{align}
\phi_m & \approx 2\pi
\biggl(
   \underbrace{\Delta f_m t}_{\text{Time}}
   - \underbrace{\Delta f_m\frac{R}{c}}_{\text{FDA beamfocusing}}
   + \underbrace{\frac{x_m\sin\theta}{\lambda}}_{\text{Beamsteering}}
   - \underbrace{\frac{x_m^{2}\cos^{2}\theta}{2R\lambda}}_{\text{Curvature}}
   \biggr), \nonumber
\label{eq:phase}
\end{align}
where $t$ denotes time and $x_m$ is the element position relative to the array center. Setting $\Delta f_m = 0$ leaves only the beamsteering and curvature terms and reduces the \ac{FDA} to a
conventional \ac{PA}. In the \ac{NF}, wavefront curvature enables beamfocusing, whereas in the \ac{FF}, the curvature vanishes, leaving only angular beamsteering.

\begin{figure*}[!t]\centering
\begingroup
\newlength{\FDATwoPanelHeight}
\setlength{\FDATwoPanelHeight}{0.25\textwidth}
\newcommand{\FDATwoPanel}[2]{%
  \IfFileExists{#1}{%
    \includegraphics[height=\FDATwoPanelHeight]{#1}%
  }{%
    \parbox[c][\FDATwoPanelHeight][c]{1.06\FDATwoPanelHeight}{%
      \centering\footnotesize\textbf{#2}\\[5pt]
      External figure asset\\not available in this preview.}%
  }%
}
\resizebox{\textwidth}{!}{%
\begin{tikzpicture}
  \matrix (fdapanels) [
    matrix of nodes,
    ampersand replacement=\&,
    nodes={inner sep=0pt,outer sep=0pt},
    column sep=5pt,
    row sep=8pt
  ] {
    \FDATwoPanel{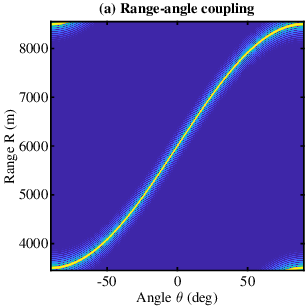}{(a) Range--angle coupling} \&
    \FDATwoPanel{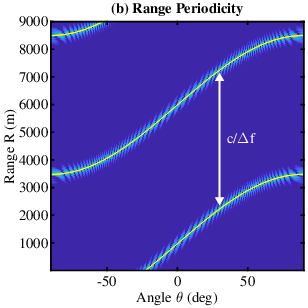}{(b) Range periodicity} \&
    \FDATwoPanel{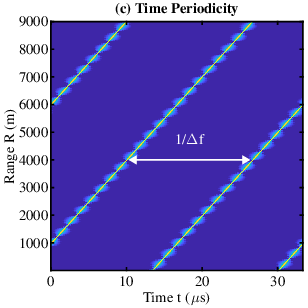}{(c) Time periodicity} \\
    \FDATwoPanel{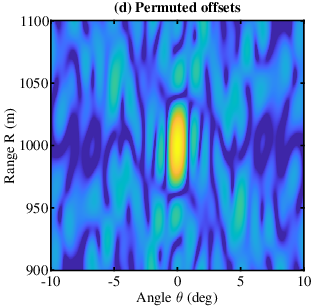}{(d) Permuted offsets} \&
    \FDATwoPanel{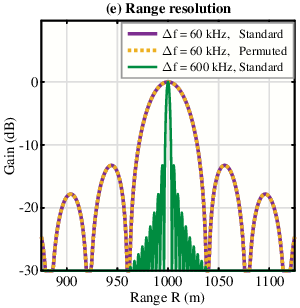}{(e) Range resolution} \&
    \FDATwoPanel{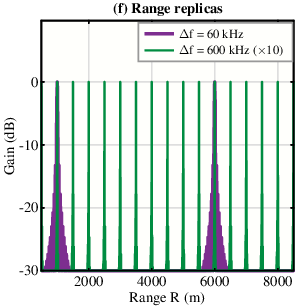}{(f) Range replicas} \\
  };
  \node[anchor=west,inner sep=0pt,outer sep=0pt]
    (fdacolorbar) at ([xshift=5pt,yshift=4pt]fdapanels.east) {%
      \IfFileExists{EPS/fig2_colorbar_long.eps}{%
        \includegraphics[height=1.88\FDATwoPanelHeight]
          {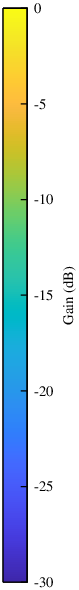}%
      }{%
        \parbox[c][1.88\FDATwoPanelHeight][c]{0.10\textwidth}{%
          \centering\footnotesize Shared\\colorbar\\(dB)}%
      }%
    };
  \begin{pgfonlayer}{background}
    \node[
      fit=(fdapanels)(fdacolorbar),
      rounded corners=5pt,
      fill=black!3,
      draw=black!20,
      line width=0.4pt,
      inner sep=6pt
    ] {};
  \end{pgfonlayer}
\end{tikzpicture}%
}
\endgroup
\caption{\ac{FDA} beampattern top row, standard offsets: (a)~range--angle at $t=0$, (b)~range--angle at $t=0$ with a larger range span, (c)~range--time at $\theta=0^\circ$; Bottom row:
  (d)~range--angle for the random-permutation design; (e)~gain versus range across the mainlobe at $\Delta f$ and $10\Delta f$; (f)~the same cut over the full span.
}
  \label{fig:beampattern}
\end{figure*}

Figure~\ref{fig:beampattern} illustrates the coupling and periodicities produced by the first three terms of the phase equation above. The example uses a 128-element \ac{ULA} operating at $f_c=28$\,GHz and focused at $(R_0,\theta_0)=(1000~\mathrm{m},0^\circ)$ with the linear profile $\Delta f_m=m\Delta f$ and $\Delta f=60$\,kHz.

\paragraph*{a) Range--Angle Coupling} In case of standard linear \ac{FDA}, the beamfocusing and the beamsteering terms are both linear in the element index. Consequently, an infinite number of range and angle combinations produce the same phase signature across the array.
The maximum gain is therefore distributed along the S-shaped range--angle ridge in Fig.~\ref{fig:beampattern}(a), rather than confined to a unique point.

\paragraph*{b) Range Periodicity} As the observation range changes, the beamfocusing term causes the element contributions to move out of phase and subsequently realign. Complete realignment occurs whenever the range changes by $c/\Delta f$, so the entire range--angle ridge repeats after this interval. For the parameters above, the period is $c/\Delta f=5000$\,m. Thus, the ridge passing through $(1000~\mathrm{m},0^\circ)$ has a replica passing through $(6000~\mathrm{m},0^\circ)$, producing the two S-shaped maximum-gain loci in Fig.~\ref{fig:beampattern}(b).

\paragraph*{c) Time Periodicity} The first term $m\Delta f t$ evolves linearly with time and an element-dependent phase-rotation rate proportional to $m$. Unlike the stochastic time variation caused by fading or physical Doppler, this programmed phase evolution causes the beampattern to periodically auto-scan across the range domain. At a fixed angle, the maximum-gain ridge moves outward at the speed of light and repeats every $1/\Delta f=16.67$\,$\mu$s, producing the recurring trajectories shown in Fig.~\ref{fig:beampattern}(c).

These replicas differ from \ac{PA} grating lobes. Grating lobes arise from spatial undersampling, typically when the element spacing exceeds half a wavelength. FDA periodicity instead arises from the frequency-offset profile and therefore persists even with half-wavelength spacing.

\subsection*{\hspace{7 pt} \textbf{Mitigating Coupling and Periodicities}}
The effects shown in Fig.~\ref{fig:beampattern} lead to distinct communication impairments. The range--angle ridge illuminates unintended locations, range replicas introduce spatial ambiguity, and temporal scanning produces intra-symbol channel variation. Each therefore requires a distinct mitigation strategy.

\paragraph*{a) Range--Angle Coupling} A localized focal spot can be synthesized using nonuniform frequency-offset profiles, such as logarithmic assignments, but arbitrary frequency values generally do not lie on the discrete \ac{OFDM} subcarrier grid. A practical alternative is to use linear frequency offsets that align with \ac{OFDM} grid but randomly permute them across the array~\cite{11162170}. The permutation disrupts the element-wise linear progression responsible for the ridge (range-angle coupling) while preserving grid-aligned frequencies. It consequently transforms Fig.~\ref{fig:beampattern}(a) into the localized spot in Fig.~\ref{fig:beampattern}(d). Because the span of frequency offsets remains unchanged, the \ac{FF} beamdepth, governed by $M\Delta f$, is also preserved, as shown in Fig.~\ref{fig:beampattern}(e).

\paragraph*{b) Range Replicas} The periodicity in range can be removed by using a nonuniform frequency-offset profile. For linear frequency offset profiles, range replicas can instead be placed outside the coverage region by selecting $\Delta f\leq c/R_{\max}$, where $R_{\max}$ is the maximum operating range of the \ac{BS}. {Figure~\ref{fig:beampattern}(f) shows the resulting trade-off. Increasing the increment from $60$ to $600$\,kHz narrows the range response by a factor of ten, but also reduces the replica spacing from $5000$ to $500$\,m. Thus, a larger offset span improves range resolution at the cost of a smaller ambiguity-free range.}

\paragraph*{c) Time Variance} The scan observed in Fig.~\ref{fig:beampattern}(c) cannot be eliminated by attempting to synthesize a strictly time-invariant physical FDA beampattern~\cite{8631015}. In radar, its impact on time on target can be mitigated through wideband pulse-compression waveforms, such as linear frequency-modulated chirps~\cite{9057739}. In communications, a more useful remedy is to constrain the frequency offset to an integer multiple of the \ac{OFDM} subcarrier spacing. The receive \ac{FFT}, evaluated over the useful symbol duration $T_u=1/\delta f$, then spans an integer number of offset cycles and orthogonally separates the shifted components. Under ideal synchronization, this avoids offset-induced \ac{ICI} and makes the post-demodulation channel on each subcarrier time-invariant, even though the physical beampattern continues to scan. The next section develops this \ac{FDA}--\ac{OFDM} integration in detail.

\begin{figure*}[!t]\centering
\FDABackground{\resizebox{\dimexpr\textwidth-12.4pt\relax}{!}{\input{fig_ofdm}}}
\caption{FDA-OFDM architectures with frequency offsets applied across the antenna elements. $s_n$ denotes the data symbol on baseband subcarrier $n$ and bin $k$ denotes the receive FFT bin.}
\label{fig:fda_ofdm_arch}
\end{figure*}
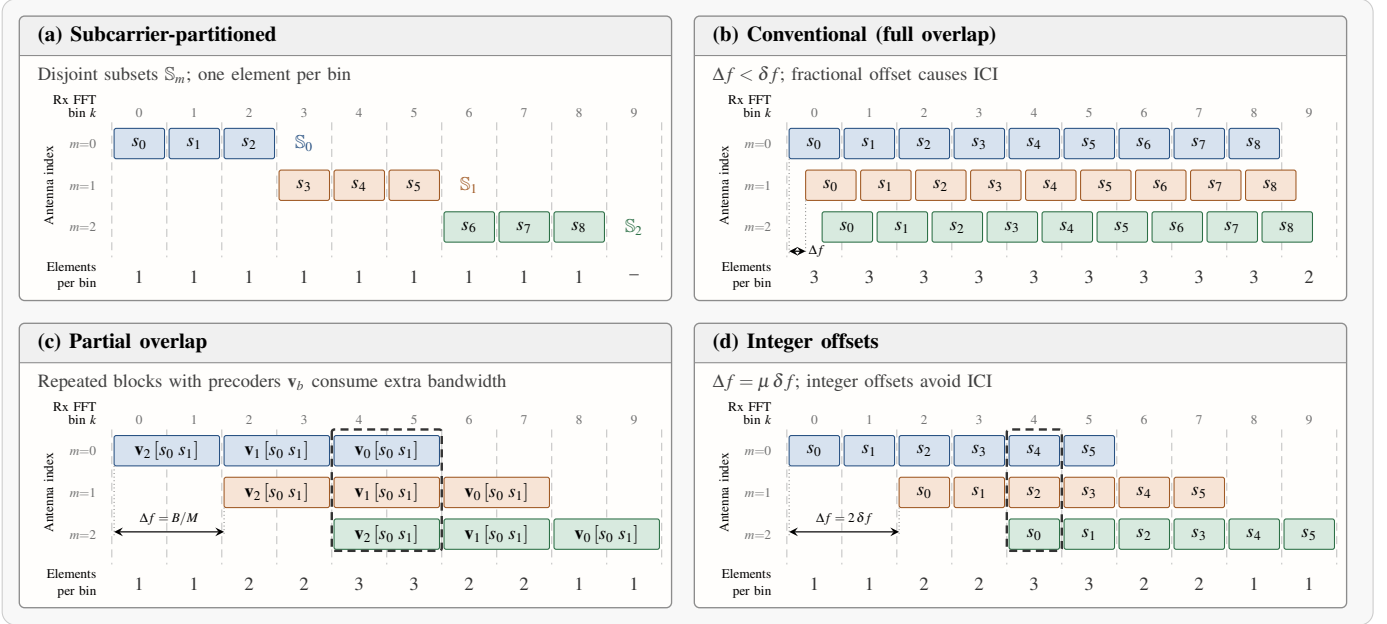

\section*{\textbf{Coexistence of FDA and OFDM}}
\label{sec:OFDM}
Since \ac{OFDM} underpins contemporary wireless systems, compatibility with this waveform is essential to the practical adoption of \acp{FDA}. Understanding how antenna-level frequency offsets interact with the \ac{OFDM} subcarrier grid is therefore crucial to assessing whether FDA range selectivity can be exploited while preserving reliable data transmission.

\subsection*{\hspace{7 pt} \textbf{FDA--OFDM Architectures}}
An \ac{OFDM} receiver uses an \ac{FFT} to separate signals on equally spaced subcarriers. Each receive bin corresponds to one frequency on this grid, whose spacing is denoted by $\delta f$. Applying a frequency offset at an antenna shifts its entire subcarrier grid. The four panels of Fig.~\ref{fig:fda_ofdm_arch} illustrate how different subcarrier assignments and frequency offset profiles affect orthogonality, array gain, and bandwidth use.

\paragraph*{a) Subcarrier Partitioned} Each antenna transmits a distinct subset of subcarriers as shown in Fig.~\ref{fig:fda_ofdm_arch}(a), where the three antennas occupy separate frequency blocks, so each receive bin contains a symbol from only one antenna. This avoids overlap between antennas, but each symbol is transmitted by a single element and therefore receives no coherent transmit array gain. For physical-layer security, the subcarrier-to-antenna assignment can be changed from symbol to symbol while compensating the transmit phases toward the intended receiver. This preserves the desired symbol constellations in that direction while introducing varying phase distortions at eavesdroppers in other directions, making their data recovery more difficult~\cite{ding2015frequency}.

\paragraph*{b) Conventional (full overlap)} All antennas transmit the same \ac{OFDM} waveform with small frequency offsets, as shown in Fig.~\ref{fig:fda_ofdm_arch}(b). When an offset is a fraction of the subcarrier spacing, the shifted tones no longer align with the receiver grid. Their energy leaks into neighboring \ac{FFT} bins, which results in offset-induced \ac{ICI}. As a result, conventional per-subcarrier equalization cannot directly accommodate the resulting intra-symbol variation.

\paragraph*{c) Partial Overlap} Repeating data blocks across frequency allows copies of the same symbols from different antennas to overlap on selected subcarriers~\cite{wachowiak2026frequency}. In the highlighted block of Fig.~\ref{fig:fda_ofdm_arch}(c), all three antennas contribute the same two symbols with different precoding weights. These contributions can combine coherently at each shared frequency, restoring array gain without frequency-offset beating between the co-frequency copies. In the \ac{FF}, this coherent combination provides angular steering as in a \ac{PA}; it does not retain the \ac{FDA} range discrimination associated with combining different carrier frequencies. The repeated blocks also consume spectral resources that could otherwise carry new data.

\paragraph*{d) Integer Offsets} Each antenna shifts its subcarrier grid by an integer number of spacings. For example, Fig.~\ref{fig:fda_ofdm_arch}(d) uses $\Delta f=2\delta f$ for shifting successive antenna spectra by two bins. The tones remain aligned with the receive grid, preserving orthogonality and avoiding offset-induced \ac{ICI} under ideal synchronization. Unlike panel (c), however, the highlighted bin contains different symbols: $s_4$, $s_2$, and $s_0$. Thus, frequency alignment alone does not separate all antenna contributions; the symbol mapping, precoding, and receiver processing must account for their overlap. Copies of the same symbol occupy different frequencies and retain the frequency-dependent propagation phases that can be used for range-selective transmission. This architecture therefore provides the basis for the \ac{FDA}-\ac{LDMA} comparison developed in the next section. Integer offsets preserve compatibility with the \ac{OFDM} grid, but introduce two practical costs:
\begin{itemize}
\item \textit{Data Bandwidth Contraction:} For a fixed total bandwidth $B$, the offset span $B_s=(M-1)\Delta f$ leaves $B-B_s$ for the common data waveform. Larger offsets improve range selectivity but reduce the bandwidth available for data transmission.

\item \textit{Edge Subcarrier Gain-Loss:} Fewer antennas contribute to the outermost receive bins, as indicated by the \emph{elements per bin} counts in Fig.~\ref{fig:fda_ofdm_arch}(d). These edge bins therefore cannot provide the same array gain as bins shared by all antennas.
\end{itemize}

\subsection*{\hspace{7 pt} \textbf{FDA--OFDM for ISAC}}
Originally introduced for radar sensing~\cite{1631800}, \acp{FDA} are also relevant to future wireless networks that are expected to offer \ac{ISAC}. The central question is therefore how \ac{FDA} range selectivity can complement \ac{OFDM} waveform-based sensing while preserving data transmission.

\begin{figure*}[!t]\centering
\begingroup
\newcommand{\FDAFourPanel}[2]{%
  \IfFileExists{#1}{%
    \includegraphics[height=0.29\textwidth]{#1}%
  }{%
    \parbox[c][0.29\textwidth][c]{0.29\textwidth}{%
      \centering\footnotesize\textbf{#2}\\[5pt]
      External figure asset\\not available in this preview.}%
  }%
}
\resizebox{\textwidth}{!}{%
\begin{tikzpicture}
  \matrix (fdafourpanels) [
    matrix of nodes,
    ampersand replacement=\&,
    nodes={inner sep=0pt,outer sep=0pt},
    column sep=8pt
  ] {
    \FDAFourPanel{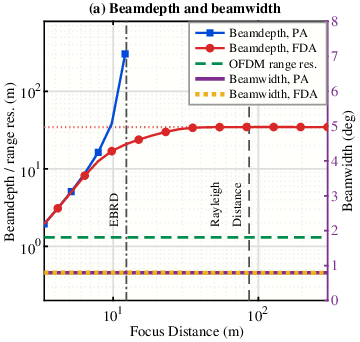}{(a) Beamdepth and beamwidth} \&
    \FDAFourPanel{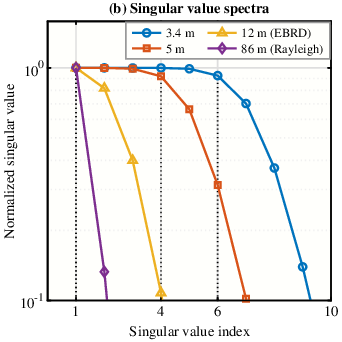}{(b) Singular-value spectra} \&
    \FDAFourPanel{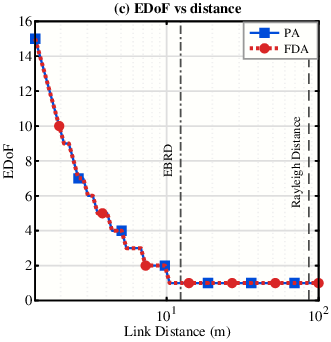}{(c) Effective degrees of freedom} \\
  };
  \begin{pgfonlayer}{background}
    \node[fit=(fdafourpanels),rounded corners=5pt,
      fill=black!3,draw=black!20,line width=0.4pt,inner sep=6pt] {};
  \end{pgfonlayer}
\end{tikzpicture}%
}
\endgroup
\caption{
Focal geometry and the spatial degrees of freedom: (a)~beamdepth and the \ac{OFDM} range resolution (left) and beamwidth (right) vs. focus distance,
 (b)~the
spectra vs. singular value index at four link distances, (c)~the \acs{EDoF} vs. link distance.
}
\label{fig:dof}
\end{figure*}

\paragraph*{a) OFDM Radar Fundamentals} In monostatic sensing, where the transceivers are co-located, the receiver processes echoes using the known transmitted data. The phase progression across subcarriers reveals the round-trip propagation delay and hence the target range. On the other hand, the phase evolution across successive symbols reveals Doppler and thus radial velocity.
{For conventional radar processing over a contiguous bandwidth $B$, the nominal range resolution is $c/(2B)$~\cite{11095387}, independent of the target distance. Since the offset span $B_s$ is smaller than $B$, the radar range resolution remains below the beamdepths of both the \ac{PA} and \ac{FDA} across the focal distances considered in Fig.~\ref{fig:dof}(a).}
In contrast, the subcarrier spacing sets the delay-ambiguity interval: using every subcarrier on the uniform grid gives an unambiguous range interval of $c/(2\delta f)$. These quantities answer different questions: \textit{bandwidth determines how closely spaced targets can be distinguished, whereas subcarrier spacing determines when delay signatures repeat.}

Conventional \ac{FFT}-based radar processing also requires the echo delay to fit within the \ac{CP}, measured relative to the receiver's timing reference, to avoid \ac{ISI}. A \ac{CP} sufficient for the communication channel's multipath delay spread may therefore be insufficient for the desired sensing range. Note that this is a processing constraint, not a fundamental range-ambiguity limit. Meanwhile, choosing $\delta f$ large compared with the expected Doppler shift reduces Doppler-induced \ac{ICI}.{ With an unchanged \ac{CP} and one Doppler sample per symbol, increasing $\delta f$ shortens the symbol interval and expands the unambiguous velocity interval, but reduces the unambiguous range. %
Finer velocity resolution instead requires a longer coherent observation.}

\paragraph*{b) Complementary FDA Range Selectivity} Since \ac{OFDM} radar separates echoes by propagation delay, an \ac{FDA} additionally makes the array response range dependent through its antenna-level frequency offsets. Its transmit beamdepth describes spatial selectivity at a given time and angle, rather than the separation between two delay-resolved radar targets. The two measures should therefore not be equated. When $B_s$ is much smaller than the processed waveform bandwidth, waveform-based ranging can distinguish targets lying within the same broader \ac{FDA} range-selective region. A similar distinction applies to \ac{NF}-\ac{PA} beamdepth~\cite{11095387}; there is no universal resolution ordering independent of bandwidth, aperture, and focal range.

With suitable waveform separation and joint range--angle processing, the range-dependent \ac{FDA} response can help discriminate a target from clutter at other ranges, including clutter arriving from the same direction~\cite{sajid2026}. This benefit depends on the offset profile and receiver design, as range dependence alone does not guarantee clutter suppression or remove the periodic ambiguities discussed earlier. For communication, the same spatial selectivity supports the \ac{LDMA} operation examined next.

\section*{\textbf{From Range Selectivity to Spatial Multiplexing}}\label{sec:dof}
Having established how \acp{FDA} can coexist with \ac{OFDM}, we now examine when their range selectivity translates into communication gains. Two distinct questions arise: can it support additional independent data streams within a single point-to-point link, and can it help separate multiple users that are close in angle but differ in range?

\subsection*{\hspace{7pt} \textbf{Single-User \acs{MIMO}}}
\label{sec:single_user_mimo}
Spatial \ac{DoF} represent the number of independent subchannels that can support parallel communication modes and, hence, multiple data streams. They are quantified by the singular values of the \ac{MIMO} channel.
For this comparison, modes with singular values above a prescribed threshold are counted toward the \ac{EDoF}.

From an electromagnetic perspective, the \ac{EDoF} represents the number of distinct transmit beams that can be spatially resolved by the receive aperture along a given spatial dimension, such as angle or distance. %
Accordingly, angular-domain and distance-domain multiplexing are fundamentally related to the beamwidth and beamdepth, respectively.

\paragraph*{a) Angular Domain Multiplexing}
Since beamwidth is determined solely by the aperture length in wavelengths, the \ac{PA} and \ac{FDA} exhibit almost identical beamwidths. As demonstrated in Fig.~\ref{fig:dof}(a) for a $128$-element \ac{ULA} operating at $28$\,GHz, both arrays have the same beamwidth for a given aperture. In the \ac{FF}, this results in an identical and constant angular \ac{EDoF} under \ac{LoS} conditions. As the \ac{UE} moves closer to the array and enters the \ac{NF}, however, the same angular beamwidth corresponds to a narrower physical footprint. Together with the spatial variation introduced by spherical wavefronts, this enables finer spatial resolution and, consequently, a substantial increase in the angular \ac{EDoF} in the \ac{NF}.

\paragraph*{b) Distance Domain Multiplexing}
In the \ac{FF}, the beamdepth of an \ac{FDA} is determined by the frequency offsets. In the \ac{NF}, however, its beamdepth is governed jointly by the aperture-dependent and frequency-offset-dependent beamdepths.
At short focal distances, wavefront curvature dominates the range selectivity, making the \ac{FDA} beamdepth nearly identical to that of a conventional \ac{PA} up to approximately $6$\,m, well within the \ac{EBRD}. As the focal distance increases beyond the \ac{EBRD}, the \ac{FDA} beamdepth approaches $0.886c/(M\Delta f)$, whereas the \ac{PA} beamdepth becomes unbounded [Fig.~\ref{fig:dof}(a)]. The \ac{EBRD} defines the maximum focal distance at which a \ac{PA} can maintain finite beamdepth in the \ac{NF}~\cite{ebrd}. However, exploiting distance-domain \ac{EDoF} in either the \ac{NF} or \ac{FF} requires the receive aperture to be oriented along the radial direction. Moreover, beamdepth is generally much larger than beamwidth when measured in physical distance \cite{ebrd}, implying that resolving multiple distance-domain beams requires an impractically large receive aperture. Hence, practical array dimensions cannot resolve multiple distance beams, and the distance-domain \ac{EDoF} remains limited to one in the \ac{FF}.

\begin{figure*}[!t]\centering
\begingroup
\newcommand{\FDAFivePanel}[2]{%
  \IfFileExists{#1}{%
    \includegraphics[height=0.29\textwidth]{#1}%
  }{%
    \parbox[c][0.29\textwidth][c]{0.29\textwidth}{%
      \centering\footnotesize\textbf{#2}\\[5pt]
      External figure asset\\not available in this preview.}%
  }%
}
\resizebox{\textwidth}{!}{%
\begin{tikzpicture}
  \matrix (fdafivepanels) [
    matrix of nodes,
    ampersand replacement=\&,
    nodes={inner sep=0pt,outer sep=0pt},
    column sep=8pt
  ] {
    \FDAFivePanel{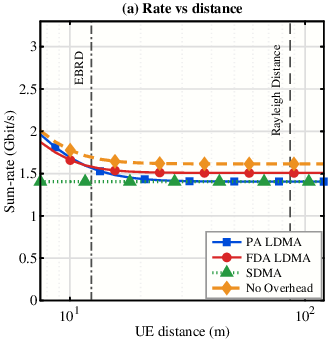}{(a) Sum-rate versus distance} \&
    \FDAFivePanel{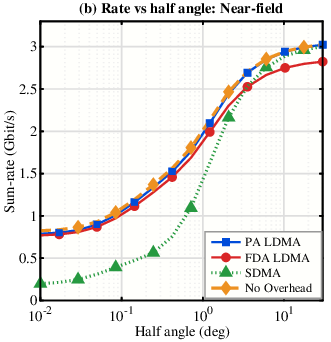}{(b) Near-field angular spread} \&
    \FDAFivePanel{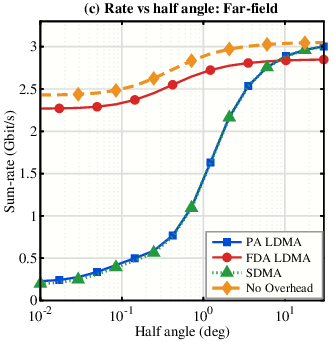}{(c) Far-field angular spread} \\
  };
  \begin{pgfonlayer}{background}
    \node[fit=(fdafivepanels),rounded corners=5pt,
      fill=black!3,draw=black!20,line width=0.4pt,inner sep=6pt] {};
  \end{pgfonlayer}
\end{tikzpicture}%
}
\endgroup
\caption{Total downlink sum-rate for SDMA, \ac{PA}-\ac{LDMA}, \ac{FDA}-\ac{LDMA} and the ``No Overhead'' ceiling. (a)~Versus UE distance,
  with user ranges drawn from a window spanning $\pm4$\,m about each distance and a sector half-angle of $1^\circ$. (b),(c)~Versus sector half-angle,
  with ranges drawn from $[3.5,\,11.5]$\,m in the \ac{NF} and $[86,\,351]$\,m in the
  \ac{FF}.}
  \label{fig:se}
\end{figure*}

\subsection*{\hspace{7pt}  \textbf{Multiuser Multiplexing}}
State-of-the-art communication systems employing \acp{PA} in the \ac{FF} exploit \ac{SDMA} to separate \acp{UE} based on their angular direction. However, \acp{UE} at similar angles experience interference and cannot be simultaneously served through \ac{SDMA}.
In contrast, the spherical wavefronts in the \ac{NF} enable beamfocusing in the joint angle-distance domain. This allows a \ac{PA} to employ \ac{LDMA} by simultaneously serving \acp{UE} located at the same angle but different distances. The \ac{LDMA} capability of a \ac{PA}, however, is limited to the \ac{EBRD} of the \ac{NF}, beyond which the beamdepth becomes unbounded. An \ac{FDA} can overcome this limitation by enabling distance-domain multiplexing in the \ac{FF}, owing to its finite beamdepth, as discussed earlier.

The reduced propagation distance in the \ac{NF} also improves angular spatial multiplexing for both \acp{PA} and \acp{FDA}. Specifically, the physical beamwidth is approximately given by $R\Delta\theta$, where $R$ denotes the radial distance and $\Delta\theta$ is the angular beamwidth in radians. Thus, for a fixed angular beamwidth, the physical beamwidth decreases at near ranges. Consequently, a sector can accommodate more resolvable beams at shorter ranges, enabling the \ac{NF} to multiplex more \acp{UE} simultaneously.

The system bandwidth $B$ is also the data bandwidth for the \ac{PA}. However, the \ac{FDA} sacrifices $B_{s}=(M-1)\Delta f$ for beamfocusing and is thereby left with a data bandwidth of $B-B_s$
(Section: Coexistence of FDA and OFDM). That is, $B_s$ is the beamfocusing cost and has a direct impact on the achievable rate of the \ac{FDA}, which is studied in the case study below.

\subsection*{\hspace{7pt} \textbf{Case Study}}
This case study compares the \ac{EDoF} and spatial multiplexing capabilities of the \ac{PA} and the \ac{FDA}.
For the single-user case in Fig.~\ref{fig:dof}, the \ac{BS} is equipped with a 128-element \ac{ULA}, while the \ac{UE} employs a 64-element \ac{ULA}. The OFDM parameters are: ${S=1920}$ subcarriers, ${\delta f=60}$\,kHz, ${B\approx 115}$\,MHz, ${\Delta f=60}$\,kHz (${B_s\approx7.6}$\,MHz) with randomly permuted offsets. For the multiuser case demonstrated in Fig.~\ref{fig:se},
the same \ac{BS} serves $K=4$ single-antenna \acp{UE} at high \ac{SNR} with \ac{MMSE} precoding. The presented results are obtained by averaging over $3000$ Monte Carlo trials.

\paragraph*{a) Single-User}
The multiplexing capability of the single-user case is characterized by the singular values of the channel matrix. The normalized singular values $\sigma_i$ of $\mathbf{H}$ are plotted at different distances in Fig.~\ref{fig:dof}(b), wherein the normalized singular values above a threshold ($\sigma_i>0.9$) represent the \ac{EDoF}. As the UE moves away from the \ac{BS}, the \ac{EDoF} count decreases and ultimately becomes one at about $10$\,m, which is close to the \ac{EBRD} boundary. Next, the \ac{EDoF} count is plotted against the link distance in Fig.~\ref{fig:dof}(c), where it can be observed that the \ac{PA}'s and \ac{FDA}'s \ac{EDoF} numbers are equal to each other both in the \ac{FF} and in the \ac{NF}, which supports the claims made in
the Single-User \acs{MIMO} subsection.

\paragraph*{b) Multiuser \acs{MIMO}}
The multiplexing capability of the multiuser case is measured by the sum-rates. In Fig.~\ref{fig:se}, the total sum-rates are compared for \ac{SDMA} and \ac{LDMA}~\cite{10123941, 11162170} with \ac{PA} and \ac{FDA}. Following~\cite{11162170}, we evaluate \ac{FDA}-\ac{LDMA} with an idealized narrowband channel model. This model assumes that each \ac{UE} keeps the \ac{BS} timing instead of synchronizing to its received signal, that the \ac{CP} covers the propagation delay, and that the frequency-shifted copies of each symbol combine coherently across \ac{FFT} bins. For benchmarking purposes, we also introduce the ``No Overhead'' case, where we hypothetically assume that bandwidth spread cost $B_s$ is free. In this way, we aim to show performance degradation caused by the bandwidth used for beamfocusing.

Fig.~\ref{fig:se}(a) presents the sum-rate as a function of distance for \acp{UE} uniformly distributed over the angular range $\theta \in[-1^\circ,+1^\circ]$. At short distances, \ac{PA}-\ac{LDMA} achieves a higher sum-rate because the \ac{PA} provides a finite beamdepth that reduces inter-user correlation without requiring additional bandwidth. Beyond the \ac{EBRD}, however, the \ac{PA} loses its ability to provide a finite beamdepth and its performance approaches that of \ac{SDMA}. In contrast, \ac{FDA}-\ac{LDMA} maintains a finite beamdepth even in the \ac{FF}. As a result, \ac{FDA} achieves a higher sum-rate in the \ac{FF}.

In Fig.~\ref{fig:se}(b) and (c), we evaluate the sum-rate as a function of the angular spread. In Fig.~\ref{fig:se}(b), the \acp{UE} are restricted to the \ac{NF} region, with user distances uniformly distributed over $R \in [3.5, 11.5]$\,m. In contrast, Fig.~\ref{fig:se}(c) considers \acp{UE} located in the \ac{FF} region, with distances distributed over $R \in [86, 351]$\,m. As shown in Fig.~\ref{fig:se}(b), operating below the \ac{EBRD} allows the \ac{PA} to generate a beamdepth comparable to that of the \ac{FDA} without incurring additional bandwidth overhead. Consequently, the \ac{PA} yields higher sum-rates than the \ac{FDA} across the entire simulated angular sweep.

Conversely, Fig.~\ref{fig:se}(c) demonstrates that the \ac{PA} loses its range selectivity in the \ac{FF}, resulting in performance equivalent to standard \ac{SDMA}. At very small angular spreads, where the \acp{UE} must be distinguished primarily in the range domain, the \ac{PA} sum-rate falls to 0.2\,Gbit/s. In contrast, \ac{FDA}-\ac{LDMA} achieves 2.3\,Gbit/s by leveraging its frequency offsets. As the angular spread increases, the performance gap between the \ac{PA} and \ac{FDA} narrows because the \ac{PA} increasingly exploits angular-domain multiplexing. Ultimately, in this configuration, the \ac{FDA} achieves higher sum-rates for angular half-spreads below $\approx 8^\circ$, which corresponds to a total sector angle of $\approx 16^\circ$. Beyond this angular threshold, the \acp{UE} are sufficiently separated in the angular domain, and the bandwidth overhead of the \ac{FDA} leads to lower sum-rates compared to the \ac{PA}.

\section*{\textbf{Conclusion and Future Directions}}

This article compared \acp{PA} and \acp{FDA} in terms of beamfocusing, \ac{OFDM} compatibility, and spatial multiplexing across the \ac{NF} and \ac{FF}. The \ac{PA} has a finite beamdepth limited to the \ac{EBRD}, whereas the \ac{FDA} extends beamfocusing into the \ac{FF} at the cost of additional bandwidth. For the configurations studied, both arrays offer approximately the same beamwidth and single-user \ac{EDoF}. The main \ac{FDA} benefit is instead multiuser separation: \ac{FDA}-\ac{LDMA} improves the \ac{FF} sum-rate in narrow angular sectors under the idealized channel model, while the offset-bandwidth cost favors the \ac{PA} in the \ac{NF} and wider sectors. Translating this benefit into practical systems motivates the following research directions.

\subsection*{\hspace{7pt} \textbf{Hybrid Architectures and Hardware Impairments}}

Hybrid architectures offer a route to lower hardware cost and power consumption. In a hybrid \ac{FDA}, frequency offsets can be applied at the \ac{RF}-chain level, such that all antenna elements within a subarray share a common frequency offset. This reduces the number of independently controlled frequency offsets, but its impact on beamdepth, range replicas, and focal-point localization remains to be characterized. Moreover, phase noise and frequency errors can perturb the frequency-dependent phase profile, potentially shifting or broadening the focal region and thereby motivating dedicated calibration methods for \acp{FDA}.

\subsection*{\hspace{7pt} \textbf{FDA-Enabled ISAC}}
An \ac{FDA} embeds range information into its transmit steering vector, thereby enabling a single waveform transmission to simultaneously support \ac{LDMA} and distance-dependent sensing. Within the \ac{ISAC} framework, the applied frequency offsets dictate both the communication bandwidth overhead and the sensing beamdepth. Consequently, jointly optimizing these offsets with standard \ac{OFDM} parameters remains a critical, unresolved challenge. In this context, index modulation that conveys additional data bits through the dynamic selection of active frequency offsets offers a promising strategy to mitigate this bandwidth overhead and elevate overall spectral efficiency \cite{10534073}. Furthermore, fusing traditional \ac{OFDM} radar range estimation with the spatial selectivity of the \ac{FDA} presents a promising mechanism to resolve the spatial ambiguities introduced by periodic range replicas \cite{11672818}. Finally, extending these stationary designs to accommodate mobile \acp{UE} and dynamic targets, where the deterministic per-element phase drift heavily couples with the physical Doppler shifts, represents a complex but essential frontier for future research.

\subsection*{\hspace{7pt} \textbf{Frequency Offset and Waveform Design}}
The achievable sum-rate in \ac{FDA}-\ac{LDMA} systems does not scale monotonically with the bandwidth allocated to the array's frequency shifts. The optimal allocation is highly dependent on the spatial distribution and the total number of \acp{UE}. Because the relative penalty of the \ac{FDA} frequency overhead diminishes as the total system bandwidth grows, adaptive bandwidth allocation emerges as an effective strategy to dynamically balance the superior distance-domain user separation of the \ac{FDA} with the high raw data throughput of the \ac{PA}. Looking forward, extending this multiplexing paradigm to planar arrays will require synthesizing multidimensional frequency-offset profiles capable of forming highly localized focal spots across range, azimuth, and elevation, while strictly adhering to the discrete \ac{OFDM} subcarrier grid \cite{sajid2026}. Finally, developing novel waveforms and symbol-to-antenna mapping schemes that mitigate edge-subcarrier gain losses, repurpose the offset bandwidth for active data transmission, and accommodate non-integer frequency offsets remain promising avenues for future research.

\subsection*{\hspace{7pt} {\textbf{Physical-Layer Security}}}
{Frequency-to-antenna assignments offer security design opportunities in \ac{FDA}--\ac{OFDM}. Permuting grid-aligned offsets reshapes off-target responses without increasing the offset span. Joint permutation, precoding, and artificial-noise design may improve secrecy rates or artificial-noise jamming efficiency. Updating permutations may also hinder interception, but this uncertainty-based benefit can disappear if the eavesdropper learns the mapping or effective channel. Secrecy-capacity claims must therefore distinguish physical-channel advantages from mapping uncertainty. Neither permutation alone nor a fixed-time beampattern establishes secrecy. Permuting unchanged occupied frequencies also provides no inherent protection against wideband jamming. Future work should compare known, partially known, and unknown permutations using the complete received waveform, accounting for channel estimation, adversary capability, and coordination overhead.}

\bibliographystyle{IEEEtran}
\bibliography{Bibliography/IEEEabrv,Bibliography/my2bib}

\vspace{10 pt}
\noindent \textbf{Arda Akyazi} is currently an undergraduate student at the Middle East Technical University, Ankara, Türkiye.

\vspace{10 pt}
\noindent  \textbf{Ahmed Hussain} is currently a Ph.D. student at King Abdullah University of Science and Technology.

\vspace{10 pt}
\noindent \textbf{Asmaa Abdallah} received a Ph.D. in electrical engineering from the American University of Beirut, Beirut, Lebanon, in 2020. She is currently a research scientist at KAUST.

\vspace{10 pt}
\noindent \textbf{Ahmed M. Eltawil} received a Ph.D. degree in electrical engineering from the University of California, Los Angeles, CA, USA, in 2003. He is currently a full professor at KAUST.

\vspace{10 pt}
\noindent \textbf{Abdulkadir Celik} received a Ph.D. in double majors of electrical engineering and computer engineering from Iowa State University, Ames, IA, USA, in 2016. He is currently an associate professor at the University of Southampton.
\vspace{10 pt}

\end{document}

%% file: common.tex
\usepackage{cite}
\usepackage{amsmath,amssymb,amsfonts}
\usepackage{hyperref}
\usepackage{graphicx}
\usepackage{textcomp}
\usepackage{bm}
\usepackage{relsize}
\usepackage{amsmath}
\usepackage{amsmath, amsthm, amssymb}
\usepackage{float}
\usepackage{graphicx}
\usepackage{epstopdf}
\usepackage{lipsum}
\usepackage{caption}
\usepackage{algorithm}
\usepackage{algpseudocode}
\usepackage{verbatim}
\usepackage{calligra}
\usepackage[font={footnotesize}]{caption}
\usepackage{subfigure}
\usepackage{xcolor}
\usepackage{mathtools}
\usepackage{stackengine}
 \usepackage{booktabs,tabularx,array,amsmath}
\usepackage[scr=boondoxo,
scrscaled=1.15,]{mathalfa}
\usepackage{bbm}
\usepackage{adforn}

\usepackage[T1]{fontenc}
\usepackage[utf8]{inputenc}
\usepackage{mathptmx}
\usepackage{amsmath,amssymb}
\usepackage{graphicx}
\usepackage{booktabs}
\usepackage{multirow}
\usepackage{xcolor}
\usepackage{tikz}
\usepackage{pgfplots}
\usepackage{url}
\usepackage{microtype}
\pgfplotsset{compat=1.18}
\usetikzlibrary{arrows.meta,positioning,calc,fit,backgrounds,shapes.geometric,
                decorations.pathreplacing,decorations.markings,patterns,shadows.blur}
\usepgfplotslibrary{groupplots,fillbetween}

\definecolor{cA}{HTML}{08519C}   %
\definecolor{cB}{HTML}{CB2027}   %
\definecolor{cC}{HTML}{E08214}   %
\definecolor{cD}{HTML}{1B7837}   %
\definecolor{cE}{HTML}{6A3D9A}   %
\definecolor{cF}{HTML}{2B8CBE}   %
\definecolor{ink}{HTML}{1A1A1A}
\definecolor{gr}{HTML}{6E7B85}
\definecolor{fillA}{HTML}{DCE9F5}
\definecolor{fillB}{HTML}{FBE3E2}
\definecolor{fillC}{HTML}{FDEBD2}
\definecolor{fillD}{HTML}{DDF0E3}
\definecolor{fillE}{HTML}{EAE1F3}
\definecolor{fillG}{HTML}{EEF1F3}

\definecolor{e0c}{RGB}{55,110,175}
\definecolor{e1c}{RGB}{214,120,50}
\definecolor{e2c}{RGB}{60,150,95}
\pgfplotsset{
  every axis/.append style={
    line width=0.45pt,
    axis line style={draw=ink!70,line width=0.45pt},
    tick style={draw=ink!70,line width=0.4pt},
    label style={font=\scriptsize,inner sep=1pt},
    tick label style={font=\tiny},
    title style={font=\scriptsize\bfseries,yshift=-2pt},
    legend style={font=\tiny,draw=ink!35,fill=white,fill opacity=0.9,
                  text opacity=1,inner sep=1.5pt,row sep=-1pt,rounded corners=1pt},
    grid style={draw=ink!12,line width=0.3pt},
    major tick length=1.6pt, minor tick length=0.9pt,
    scale only axis,
  },
  lp/.style={line width=0.85pt},
  bpaxis/.style={
    enlargelimits=false, axis on top, scale only axis,
    colormap/viridis, point meta min=-30, point meta max=0,
    tick align=outside, xtick pos=left, ytick pos=left,
  },
}

\tikzset{
  >={Stealth[length=3.6pt,width=2.6pt]},
  sig/.style={draw=ink!75,line width=0.45pt,->},
  sigt/.style={draw=ink!75,line width=0.45pt},
  blk/.style 2 args={draw=ink!55,fill=#1,rounded corners=1.1pt,
      minimum height=4.6mm,minimum width=#2,inner xsep=2pt,inner ysep=1pt,
      font=\tiny,align=center,line width=0.45pt},
  bx/.style={blk={fillG}{7mm}},
  bxa/.style={blk={fillA}{7mm}},
  bxb/.style={blk={fillB}{7mm}},
  bxc/.style={blk={fillC}{7mm}},
  bxd/.style={blk={fillD}{7mm}},
  bxe/.style={blk={fillE}{7mm}},
  mix/.style={draw=ink!60,circle,inner sep=0pt,minimum size=3.4mm,
      fill=white,line width=0.45pt,font=\tiny},
  amp/.style={draw=ink!60,fill=fillG,regular polygon,regular polygon sides=3,
      shape border rotate=-90,inner sep=0pt,minimum size=3.6mm,line width=0.4pt},
  ant/.style={draw=cA,line width=0.7pt},
  lbl/.style={font=\tiny,text=ink},
  slbl/.style={font=\tiny\itshape,text=gr},
  pan/.style={font=\scriptsize\bfseries,text=ink},
  note/.style={draw=ink!25,fill=white,rounded corners=1.2pt,
      font=\tiny,align=left,inner sep=2.2pt,line width=0.4pt},
  frame/.style={draw=ink!30,rounded corners=2pt,line width=0.5pt},
}
\newcommand{\antenna}[2]{%
  \draw[ant] (#1) -- ++(0,2.2mm);
  \draw[ant] ($(#1)+(0,2.2mm)$) -- ++(-1.5mm,1.6mm);
  \draw[ant] ($(#1)+(0,2.2mm)$) -- ++(1.5mm,1.6mm);
  \node[lbl,anchor=west] at ($(#1)+(1.8mm,2.9mm)$) {#2};}

%% file: fig1_combined.tex
\begin{tikzpicture}[font=\tiny,line cap=round]

\tikzset{
  pfr/.style={draw=ink!20,fill=panelbg,rounded corners=1.8pt,line width=0.4pt},
  tall/.style={draw=ink!55,fill=fillA,rounded corners=1.1pt,inner sep=1pt,
               line width=0.45pt,font=\tiny},
  hdr/.style={font=\tiny,text=gr},
  comb/.style={draw=cC,line width=0.5pt},
}
\newcommand{\pamp}[2]{%
  \node[inner sep=0pt,minimum width=4.0mm,minimum height=4.2mm] (#1) at (#2) {};
  \draw[draw=ink!60,fill=ink!10,line width=0.45pt,line join=round]
    ($(#2)+(-0.19,-0.21)$)--($(#2)+(0.21,0)$)--($(#2)+(-0.19,0.21)$)--cycle;}
\newcommand{\vdts}[1]{%
  \foreach \vk in {-0.085,0,0.085}{\fill[ink!80] ($(#1)+(0,\vk)$) circle (0.015);}}
\renewcommand{\antenna}[2]{%
  \draw[draw=cA,line width=0.5pt] (#1) -- ($(#1)+(0,0.14)$);
  \draw[draw=cA,fill=cA!25,line width=0.5pt,line join=round]
    ($(#1)+(-0.15,0.34)$) -- ($(#1)+(0,0.14)$) -- ($(#1)+(0.15,0.34)$) -- cycle;
  \node[lbl,anchor=center] at ($(#1)+(0.50,0)$) {#2};}

\draw[pfr] (0,4.75) rectangle (8.20,8.85);
\draw[pfr] (0,0)    rectangle (8.20,4.55);

\begin{scope}[shift={(0,4.50)}]
  \node[pan,anchor=north west] at (0.10,4.27) {(a) PA};

  \node[tall,minimum width=4.6mm,minimum height=31mm] (bb) at (0.46,2.10)
        {\rotatebox{90}{Baseband processing}};

  \foreach \y in {3.30,2.70,1.30,0.70}{
    \node[bx,minimum width=8.4mm,minimum height=3mm,inner sep=0pt] (rf) at (1.46,\y) {};
    \node[inner sep=0pt,anchor=base] at (1.46,\y-0.058) {RF chain};
    \pamp{am}{2.42,\y}
    \draw[sig,-] (0.69,\y) -- (rf);
    \draw[sig,-] (rf) -- (am);
    \draw[sigt] (am) -- (2.98,\y);
    \antenna{2.98,\y}{}}
  \foreach \x in {1.46,2.42}{\vdts{\x,2.00}}
  \draw[sigt] (2.48,2.00) -- (2.98,2.00);
  \antenna{2.98,2.00}{}

  \node[lbl,anchor=south,inner sep=0.5pt] at (2.98,3.682) {$\frac{M-1}{2}$};
  \node[lbl,anchor=east,inner sep=0.5pt]  at (2.84,2.24) {$0$};
  \node[lbl,anchor=east,inner sep=0.5pt]  at (2.84,1.54) {$m$};
  \node[lbl,anchor=north,inner sep=0.5pt] at (2.92,0.660) {$-\frac{M-1}{2}$};

  \node[hdr,anchor=north] at (2.32,4.10) {Power};
  \node[hdr,anchor=north] at (2.32,3.92) {Amplifier};
  \node[hdr,anchor=north] at (2.98,4.30) {Antenna};      %

  \begin{scope}
    \clip (3.40,3.409) -- (7.15,2.24) -- (3.40,1.071) -- cycle;
    \clip (3.60,0) rectangle (6.30,4.35);
    \foreach \xk in {3.90,4.32,4.74,5.16,5.58,6.00}{
      \draw[draw=cA!55,line width=0.42pt]
        ({2*\xk-7.15+(7.15-\xk)*cos(25)},{2.24+(7.15-\xk)*sin(25)})
        arc[start angle=25,end angle=-25,radius={7.15-\xk}];}
  \end{scope}
  \draw[cD,fill=cD!14,line width=0.7pt] (7.15,2.24) ellipse (0.45 and 0.13);
  \draw[draw=gr!65,line width=0.3pt,dash pattern=on 1pt off 1pt]
        (3.40,3.409) -- (7.15,2.24) -- (3.40,1.071);
  \node[slbl,text=cA,anchor=south west] at (4.45,3.12) {Spherical wavefronts};

  \draw[draw=gr!80,line width=0.3pt]
        (3.16,3.54) -- (3.32,3.54)  (3.16,0.94) -- (3.32,0.94);
  \draw[<->,draw=gr!80,line width=0.35pt] (3.24,0.94) -- (3.24,3.54);
  \node[lbl,anchor=west] at (3.27,2.62) {$D$};

  \draw[draw=gr!80,dashed,line width=0.4pt] (3.055,2.24) -- (7.15,2.24);
  \draw[draw=gr!80,dashed,line width=0.4pt] (3.066,1.554) -- (7.15,2.24);
  \node[lbl,anchor=south]  at (4.10,2.24) {$R_0$};
  \node[lbl,anchor=center] at (4.02,1.49) {$R_m$};

  \fill[cD] (7.15,2.24) circle (0.032);
  \node[lbl,text=cD,anchor=south] at (7.15,2.52) {$(R_0,\theta_0)$};

  \draw[draw=cD!70,line width=0.3pt,dash pattern=on 1.2pt off 1.2pt]
        (6.70,2.11) -- (6.70,1.94)  (7.60,2.11) -- (7.60,1.94);
  \draw[<->,draw=cD,line width=0.55pt] (6.70,1.94) -- (7.60,1.94);
  \node[lbl,text=cD,anchor=north,align=center] at (7.15,1.90)
        {beamdepth\\\\$\approx7\lambda (\tfrac{R_0}{D})^2$};

  \draw[draw=cD!70,line width=0.3pt,dash pattern=on 1.2pt off 1.2pt]
        (7.15,2.37) -- (7.77,2.37)  (7.15,2.11) -- (7.77,2.11);
  \draw[{Stealth[length=2.2pt,width=1.8pt]}-{Stealth[length=2.2pt,width=1.8pt]},
        draw=cD,line width=0.5pt] (7.77,2.11) -- (7.77,2.37);
  \node[lbl,text=cD,rotate=90,anchor=center] at (7.95,2.24) {beamwidth};
\end{scope}

\begin{scope}
  \node[pan,anchor=north west] at (0.10,4.47) {(b) FDA};

  \node[tall,minimum width=4.6mm,minimum height=25mm] (bb2) at (0.46,2.65)
        {\rotatebox{90}{Baseband processing}};

  \foreach \y/\fl in {3.75/{$f_{\frac{M-1}{2}}$}, 3.15/{}, 2.15/{$f_m$}, 1.55/{$f_{-\frac{M-1}{2}}$}}{
    \node[bx,minimum width=8.4mm,minimum height=3mm,inner sep=0pt] (rf) at (1.46,\y) {};
    \node[inner sep=0pt,anchor=base] at (1.46,\y-0.058) {RF chain};
    \pamp{am}{2.42,\y}
    \draw[sig,-] (0.69,\y) -- (rf);
    \draw[sig,-] (rf) -- (am);
    \draw[sigt] (am) -- (2.98,\y);
    \antenna{2.98,\y}{}
    \node[lbl,anchor=west,inner sep=0.5pt] at (3.19,\y+0.24) {\fl};}
  \foreach \x in {1.46,2.42}{\vdts{\x,2.65}}
  \draw[sigt] (2.48,2.65) -- (2.98,2.65);
  \antenna{2.98,2.65}{}
  \node[lbl,anchor=west,inner sep=0.5pt] at (3.19,2.89) {$f_0$};

\end{scope}

\def\xR{4.90}
\foreach \xx in {3.92,4.00,4.08}{
  \draw[draw=cA!75,line width=0.45pt] (\xx,1.62) -- (\xx,4.10);}
\node[slbl,text=cA,rotate=90,anchor=center] at (4.27,2.86) {Planar wavefronts};

\draw[draw=cD!85,line width=0.45pt,dash pattern=on 1.2pt off 1pt]
      (\xR,0.55) -- (\xR,4.12);
\draw[draw=cC!90,line width=0.45pt,dash pattern=on 1.2pt off 1pt]
      ({\xR+2.4},0.55) -- ({\xR+2.4},4.12);

\foreach \y/\n in {3.99/11, 3.39/10, 2.89/8, 2.39/6, 1.79/5}{
  \draw[draw=ink!12,line width=0.3pt] (4.45,\y) -- (8.05,\y);
  \draw[draw=cA,line width=0.55pt,samples=601,domain=4.45:8.05,variable=\x]
    plot (\x,{\y+0.10*cos(150*\n*(\x-\xR))});
  \fill[cD] (\xR,\y+0.10) circle (0.030);
  \fill[cC] ({\xR+2.4},\y+0.10) circle (0.030);}

\draw[<->,draw=ink!60,line width=0.35pt] (\xR,1.51) -- ({\xR+2.4},1.51);
\node[lbl,fill=panelbg,inner sep=1pt] at ({\xR+1.2},1.51) {$c/\Delta f$};

\draw[draw=cD,line width=0.9pt,samples=343,domain=4.40:6.10,variable=\x]
  plot (\x,{0.55+0.78*pow(sin(450*(\x-\xR))/(6*sin(75*(\x-\xR))),2)});
\draw[draw=cC,line width=0.9pt,samples=301,domain=6.10:8.05,variable=\x]
  plot (\x,{0.55+0.78*pow(sin(450*(\x-\xR))/(6*sin(75*(\x-\xR))),2)});

\draw[draw=ink!45,line width=0.3pt,dash pattern=on 1.4pt off 1.4pt]
      ({\xR-0.28},0.94) -- ({\xR+0.30},0.94);
\draw[<->,draw=cD,line width=0.6pt] ({\xR-0.1772},0.94) -- ({\xR+0.1772},0.94);
\draw[draw=cD,line width=0.4pt] ({\xR+0.1772},0.94) -- ({\xR+0.30},1.03);
\node[lbl,text=cD,anchor=west] at ({\xR+0.34},1.07)
      {beamdepth $=\dfrac{0.886c}{M\Delta f}$};

\draw[draw=ink!45,line width=0.35pt,->] (4.40,0.55) -- (8.12,0.55);
\node[lbl,anchor=north]      at (\xR,0.47)        {$R_0$};
\node[lbl,anchor=north]      at ({\xR+2.4},0.47)  {$R_0+c/\Delta f$};
\node[lbl,anchor=north east] at (8.12,0.47)       {$R$};

\end{tikzpicture}

%% file: fig_ofdm.tex
\begin{tikzpicture}[
  font=\scriptsize,
  panelbox/.style={draw=black!45, line width=0.5pt, rounded corners=2pt},
  titlebar/.style={fill=black!6},
  ptitle/.style={font=\footnotesize\bfseries, anchor=west},
  psub/.style={font=\scriptsize, text=black!75, anchor=west},
  axlab/.style={font=\tiny, text=black},        %
  tick/.style={font=\tiny, text=black!55},      %
  binline/.style={densely dashed, black!25, line width=0.3pt},
  slot/.style={line width=0.4pt, rounded corners=0.8pt},
  cnt/.style={font=\scriptsize, text=black!80},
  hl/.style={densely dashed, line width=0.9pt, draw=black!80, rounded corners=1pt}, %
  dim/.style={stealth-stealth, line width=0.45pt},
  guide/.style={densely dotted, black!55, line width=0.35pt}
]
\providecolor{e0c}{RGB}{0,114,178}
\providecolor{e1c}{RGB}{213,94,0}
\providecolor{e2c}{RGB}{0,158,115}
\colorlet{e0}{e0c}\colorlet{e1}{e1c}\colorlet{e2}{e2c}

\def\PW{8.6}     %
\def\PH{3.75}    %
\def\GX{0.3}     %
\def\GY{0.3}     %
\def\gx{1.22}    %
\def\bw{0.725}   %
\def\yk{-1.33}   %
\def\yt{-1.40}   %
\def\yr{-1.68}   %
\def\dr{0.55}    %
\def\yb{-3.18}   %
\def\ycnt{-3.43} %
\def\xant{0.39}  %

\def\fdaslot#1#2#3#4#5{%
  \pgfmathsetmacro{\xa}{\gx+(#1)*\bw+0.03}%
  \pgfmathsetmacro{\xb}{\gx+(#1+#2)*\bw-0.03}%
  \pgfmathsetmacro{\ym}{\yr-\dr*(#3)}%
  \draw[slot, fill=#4!20, draw=#4!75!black] (\xa,\ym-0.2) rectangle (\xb,\ym+0.2);
  \node at ({(\xa+\xb)/2},\ym) {#5};}
\def\fdapanel#1#2{%
  \fill[panelbg, rounded corners=2pt] (0,0) rectangle (\PW,-\PH);
  \fill[titlebar, rounded corners=2pt] (0,0) rectangle (\PW,-0.52);
  \fill[titlebar] (0,-0.3) rectangle (\PW,-0.52);
  \draw[panelbox] (0,0) rectangle (\PW,-\PH);
  \draw[black!45, line width=0.4pt] (0,-0.52) -- (\PW,-0.52);
  \node[ptitle] at (0.12,-0.26) {#1};
  \node[psub] at (0.12,-0.78) {#2};
  \node[axlab, anchor=base east] at ({\gx-0.08},{\yk+0.17}) {Rx FFT};  %
  \node[axlab, anchor=base east] at ({\gx-0.08},\yk) {bin $k$};
  \foreach \k in {0,...,9} {\node[tick, anchor=base] at ({\gx+(\k+0.5)*\bw},\yk) {\k};}
  \foreach \k in {0,...,10} {\draw[binline] ({\gx+\k*\bw},\yt) -- ({\gx+\k*\bw},\yb);}
  \node[axlab, rotate=90] at (\xant,{\yr-\dr}) {Antenna index};
  \foreach \m in {0,1,2} {\node[tick, anchor=east] at ({\gx-0.08},{\yr-\dr*\m}) {$m{=}\m$};}
  \node[axlab, anchor=east, align=right] at ({\gx-0.08},\ycnt) {Elements\\per bin};}
\def\fdacounts#1{%
  \foreach \c [count=\k from 0] in {#1} {\node[cnt] at ({\gx+(\k+0.5)*\bw},\ycnt) {\c};}}

\begin{scope}[shift={(0,0)}]
  \fdapanel{(a) Subcarrier-partitioned}{Disjoint subsets $\mathbb{S}_m$; one element per bin}
  \foreach \k in {0,1,2} {\fdaslot{\k}{1}{0}{e0}{$s_{\k}$}}
  \foreach \k in {3,4,5} {\fdaslot{\k}{1}{1}{e1}{$s_{\k}$}}
  \foreach \k in {6,7,8} {\fdaslot{\k}{1}{2}{e2}{$s_{\k}$}}
  \node[text=e0!75!black] at ({\gx+3.5*\bw},{\yr})       {$\mathbb{S}_0$};
  \node[text=e1!75!black] at ({\gx+6.5*\bw},{\yr-\dr})   {$\mathbb{S}_1$};
  \node[text=e2!75!black] at ({\gx+9.5*\bw},{\yr-2*\dr}) {$\mathbb{S}_2$};
  \fdacounts{1,1,1,1,1,1,1,1,1,--}
\end{scope}

\begin{scope}[shift={({\PW+\GX},0)}]
  \fdapanel{(b) Conventional (full overlap)}{$\Delta f<\delta f$; fractional offset causes ICI}
  \foreach \k in {0,...,8} {\fdaslot{\k}{1}{0}{e0}{$s_{\k}$}}
  \foreach \k in {0,...,8} {\fdaslot{\k+0.3}{1}{1}{e1}{$s_{\k}$}}
  \foreach \k in {0,...,8} {\fdaslot{\k+0.6}{1}{2}{e2}{$s_{\k}$}}
  \draw[guide] ({\gx+0.03},{\yr-0.2}) -- ({\gx+0.03},{\yb+0.04});
  \draw[guide] ({\gx+0.3*\bw+0.03},{\yr-\dr-0.2}) -- ({\gx+0.3*\bw+0.03},{\yb+0.04});
  \draw[dim] ({\gx+0.03},{\yb+0.08}) -- ({\gx+0.3*\bw+0.03},{\yb+0.08})
     node[right=0.5pt, font=\tiny, inner sep=0.5pt] {$\Delta f$};
  \fdacounts{3,3,3,3,3,3,3,3,3,2}
\end{scope}

\begin{scope}[shift={(0,{-\PH-\GY})}]
  \fdapanel{(c) Partial overlap}{Repeated blocks with precoders $\mathbf{v}_b$ consume extra bandwidth}
  \foreach \m/\c in {0/e0,1/e1,2/e2} {
    \foreach \j/\v in {0/2,1/1,2/0} {
      \pgfmathtruncatemacro{\kb}{2*\m+2*\j}
      \fdaslot{\kb}{2}{\m}{\c}{$\mathbf{v}_{\v}\,[s_0\;s_1]$}
    }
  }
  \draw[hl] ({\gx+4*\bw},\yt) rectangle ({\gx+6*\bw},{\yr-2*\dr-0.22});
  \draw[guide] ({\gx+0.03},{\yr-0.2}) -- ({\gx+0.03},{\yr-2*\dr-0.02});
  \draw[guide] ({\gx+2*\bw+0.03},{\yr-\dr-0.2}) -- ({\gx+2*\bw+0.03},{\yr-2*\dr-0.02});
  \draw[dim] ({\gx+0.03},{\yr-2*\dr+0.03}) -- ({\gx+2*\bw+0.03},{\yr-2*\dr+0.03})
     node[midway, above=-1pt, font=\tiny] {$\Delta f=B/M$};
  \fdacounts{1,1,2,2,3,3,2,2,1,1}
\end{scope}

\begin{scope}[shift={({\PW+\GX},{-\PH-\GY})}]
  \fdapanel{(d) Integer offsets}{$\Delta f=\mu\,\delta f$; integer offsets avoid ICI}
  \foreach \k in {0,...,5} {\fdaslot{\k}{1}{0}{e0}{$s_{\k}$}}
  \foreach \k in {0,...,5} {\fdaslot{\k+2}{1}{1}{e1}{$s_{\k}$}}
  \foreach \k in {0,...,5} {\fdaslot{\k+4}{1}{2}{e2}{$s_{\k}$}}
  \draw[hl] ({\gx+4*\bw},\yt) rectangle ({\gx+5*\bw},{\yr-2*\dr-0.22});
  \draw[guide] ({\gx+0.03},{\yr-0.2}) -- ({\gx+0.03},{\yr-2*\dr-0.02});
  \draw[guide] ({\gx+2*\bw+0.03},{\yr-\dr-0.2}) -- ({\gx+2*\bw+0.03},{\yr-2*\dr-0.02});
  \draw[dim] ({\gx+0.03},{\yr-2*\dr+0.03}) -- ({\gx+2*\bw+0.03},{\yr-2*\dr+0.03})
     node[midway, above=-1pt, font=\tiny] {$\Delta f=2\,\delta f$};
  \fdacounts{1,1,2,2,3,3,2,2,1,1}
\end{scope}
\end{tikzpicture}